\documentclass[proof]{WileyASNA-v1}

\articletype{Original Article}%

\received{2019}
\revised{2020}
\accepted{2020}

\begin{document}

\title{The Massive Binary System 9 Sgr Revisited: New Insights into Disentangling Methods\protect\thanks{Based on observations collected at the European Southern Observatory (La Silla, Chile and Cerro Paranal, Chile).}}

\author[1,2]{Edwin A. Quintero*}

\author[1]{Philippe Eenens}

\author[3]{Gregor Rauw}

\authormark{QUINTERO \textsc{et al}}

\address[1]{\orgdiv{Departamento de Astronom\'ia}, \orgname{Universidad de Guanajuato}, \orgaddress{\state{Guanajuato}, \country{Mexico}}}

\address[2]{\orgdiv{Observatorio Astron\'omico (OAUTP)}, \orgname{Universidad Tecnol\'ogica de Pereira}, \orgaddress{\state{Risaralda}, \country{Colombia}}}

\address[3]{\orgdiv{Space Sciences, Technologies and Astrophysics Research (STAR) Institute}, \orgname{Universit\'e de Li\`ege}, \orgaddress{\state{Li\`ege}, \country{Belgium}}}

\corres{*Edwin A. Quintero, Departamento de Astronom\'ia, Universidad de Guanajuato, Apartado Postal 144, 36000 Guanajuato, GTO, M\'exico. \email{equintero@utp.edu.co}}


\abstract{Disentangling techniques are often needed to obtain the spectra of the individual components of binary or multiple systems. A thorough analysis of the shift-and-add algorithm of \citet{Marchenko1998} reveals that in many cases the line fluxes are poorly reproduced and spurious wings appear. The causes of these discrepancies are discussed and a new disentangling package, QER20, is presented which significantly reduces these errors and vastly  increases the performance. When applied to the massive binary 9 Sgr, our new code yields line fluxes which are notably different from those previously published and lead us to revise the spectral classification to slightly earlier subtypes: O\,3 V ((f +)) for the primary and  O\,5V ((f)) for the secondary. We show that with the MME98 algorithm the classification of massive stars in binaries can be off by several subtypes whilst there are no such errors when the QER20 package is used.
}




\maketitle


\section{Introduction}\label{sec:intro}

In a binary or multiple system with small Doppler shifts compared to the spectral lines widths, the spectra of the individual stellar components must be separated to have access to the profiles and line fluxes of each stellar component. Several algorithms have been developed with that aim, such as the tomographic method of~\cite{BagnuoloJr1991}, the discrete Fourier transform of~\cite{Hadrava1995} and the shift-and-add method of \citet[hereafter MME98]{Marchenko1998}. The MME98 algorithm can be used either on its own, when the radial velocity curve is already known, or in conjunction with a method to determine the Doppler shifts as described by~\cite{Gonzalez2006}. It seems however that all these algorithms suffer from some drawback, either being quite demanding in computer time or producing spurious artefacts in the reconstructed line profiles. Reviews of disentangling methods have been written by~\cite{Hadrava2009} and~\cite{Pavlovski2010}.

Ideally a disentangling algorithm should be efficient (fast and easy to implement) and reliable (reproducing exactly the spectrum of the individual stellar components). To achieve this objective the best route is to choose an algorithm known for its versatility and ease of use, and improve its reliability. A good candidate is the shift-and-add method of MME98. It has been applied successfully over a hundred times: WR141~\citep{Marchenko1998}, $\gamma^2$ Vel~\citep{Eversberg1999}, GP Cep~\citep{Demers2002}, DQ Vel~\citep{Barria2013}, WR62a~\citep{Collado2013}, WR21a~\citep{Tramper2016}, WR148~\citep{Munoz2016}, SMC AB6~\citep{Shenar2018}, among others.

However it is well documented that this algorithm does not always reproduce correctly the wings of broad spectral lines such as the Balmer lines: see e.g. the cases of HD 150136~\citep{Mahy2012}, HD 194649~\citep{Mahy2015}, HD 149404~\citep{Raucq2016}; and can even generate artefacts: see for instance HD 38282~\citep{Schnurr2009}, 9~Sgr~\citep{Rauw2012} and HD 165052~\citep{Ferrero2013}. If these discrepancies could be resolved, this algorithm would meet our goals of efficiency and reliability. 

In this paper we examine the causes of the deficiencies of the MME98 algorithm (\S~\ref{sec:analysis}) and propose a novel, highly reliable and efficient disentangling package (\S~\ref{sec:algorithm}). We then apply this new package to the binary 9~Sgr and compare our disentangled results to previous attempts (\S~\ref{sec:sgr}).

\section{Analysis of the shift-and-add Method}\label{sec:analysis}

When a disentangling algorithm is applied to observed spectra it is not possible to know if the individual spectra have been reproduced exactly, as there is no direct access to them. Hence our strategy will be to create two synthetic absorption profiles meant to represent the individual stellar components $A$ and $B$ of a binary system, shift them by their respective radial velocity and add them to simulate a composite spectrum. We then apply our algorithm to a series of such spectra and compare the output to the original profiles.

\begin{figure}
\centerline{\includegraphics[width=1\columnwidth]{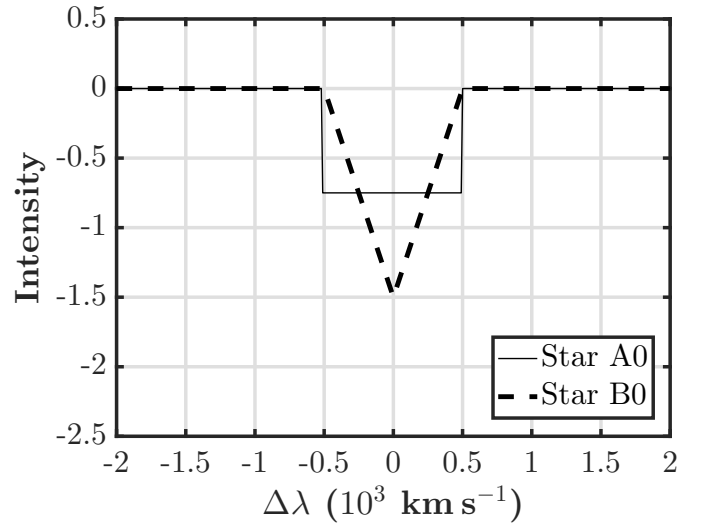}}
\caption{Synthetic line profiles of individual stellar components before blending. The rectangular shape is that of star $A$ (continuous line), the triangle of star $B$ (dotted line).\label{fig:StarA0B0}}
\end{figure}

In a first experiment we choose simple geometric shapes for our synthetic absorption profiles, so the differences between original and disentangled profiles will stand out: a rectangle having a full width half maximum (FWHM) of $1000$ km\,s$^{-1}$ for star $A$ and a triangle with FWHM of $500$ km\,s$^{-1}$ for star $B$. This ensures that the two profiles have equal line fluxes (Figure~\ref{fig:StarA0B0}). Both profiles are sampled every $10$ km\,s$^{-1}$. The continuum is set to zero. We then construct ten composite spectra by applying to both stars radial velocities ranging from $-50$ to $50$ km\,s$^{-1}$, in equal steps of $10$ km\,s$^{-1}$ (omitting the value 0). 

The first step of the first iteration of the shift-and-add method aims to reinforce the features of star $A$ in the spectrum and to minimize those of star $B$. For this, each of the $S_{k}$ composite spectra taken at times $t_{k}|^{N}_{k=1}$, is corrected for the radial velocity of star $A$ ($\upsilon_{a,k}$). They are then co-added:

\begin{eqnarray}
A_{1}(\lambda)=\frac{1}{N}\sum^{N}_{k=1}\left[S_{k}(\lambda+\upsilon_{a,k})\right]
\label{eq:starA1}
\end{eqnarray}

As a result, the flux from the line of star $B$ is spread out over a range of wavelengths and diluted. This is taken as the first approximation of star $A$.

In the second step, this approximation is subtracted from each of the composite spectra, which are then corrected for the radial velocity of $B$ ($\upsilon_{b,k}$) and co-added to provide the first approximation of star $B$:

\begin{eqnarray}
B_{1}(\lambda)=\frac{1}{N}\sum^{N}_{k=1}\left[S_{k}(\lambda+\upsilon_{b,k})-A_{1}(\lambda-\upsilon_{a,k}+\upsilon_{b,k})\right]
\label{eq:starB1}
\end{eqnarray} 

This process is repeated in the following iterations of the algorithm ($i=2,3,4,...,n$):

\begin{eqnarray}
A_{i}(\lambda)=\frac{1}{N}\sum^{N}_{k=1}\left[S_{k}(\lambda+\upsilon_{a,k})-B_{i-1}(\lambda-\upsilon_{b,k}+\upsilon_{a,k})\right]
\label{eq:starAi}
\end{eqnarray}

\begin{eqnarray}
B_{i}(\lambda)=\frac{1}{N}\sum^{N}_{k=1}\left[S_{k}(\lambda+\upsilon_{b,k})-A_{i}(\lambda-\upsilon_{a,k}+\upsilon_{b,k})\right]
\label{eq:starBi}
\end{eqnarray}

\begin{figure*}
\centerline{\includegraphics[width=2\columnwidth]{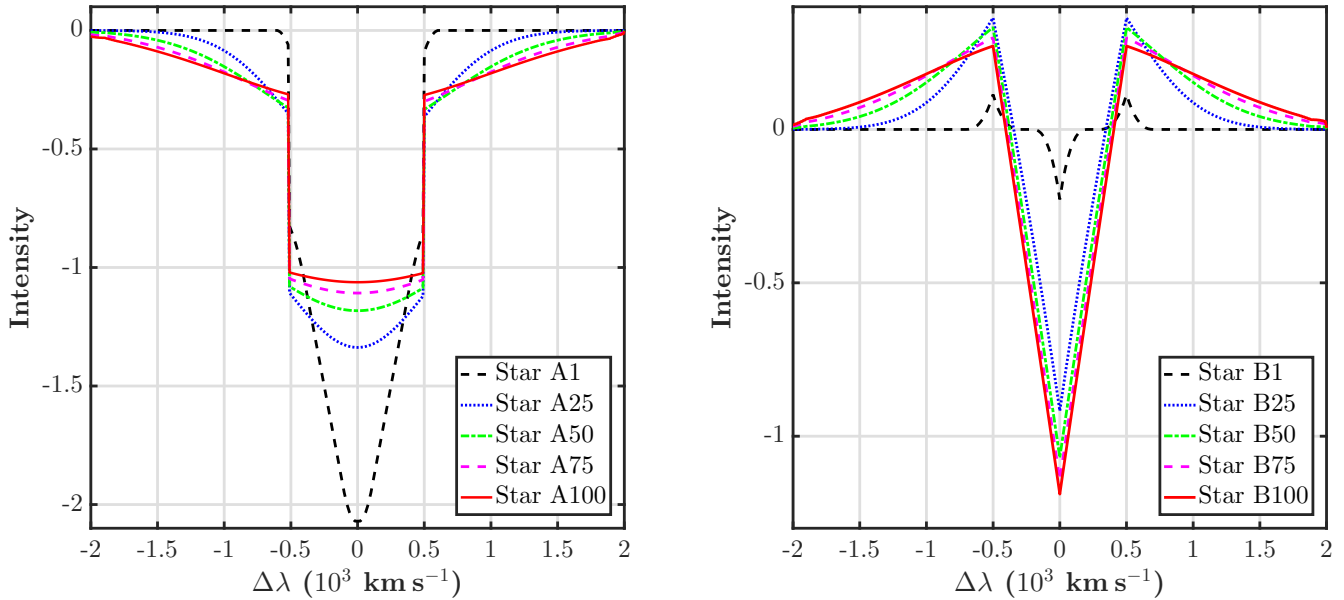}}
\caption{Output of the MME98 algorithm 
for the input shown in Figure~\ref{fig:StarA0B0}, after one, 25, 50, 75 and 100 iterations. The last two are very similar and could be interpreted wrongly as nearing  the end of the convergence process. 
Left: Star $A$. Right: Star $B$.
\label{fig:Star_1_20_40_60_80_100_200_2000}}
\end{figure*}

In Figure~\ref{fig:Star_1_20_40_60_80_100_200_2000} we can see that already at the first iteration ($i=1$) the output of the shift-and-add algorithm suffers from two kind of distortions with respect to the original, true profiles: an increase in the line flux of star $A$ and a decrease in that of star $B$; spurious absorption wings for star $A$ and spurious emission wings for star $B$. These wings are symmetrically located on each side of the absorption feature, both roughly at the same distance from the rest wavelength. This characteristic is conserved throughout the iterations. As illustrated in Figure~\ref{fig:Star_1_20_40_60_80_100_200_2000}, the flux of these wings increases during the first few iterations, and then slowly begins to decrease as it is spread out further and further along the velocity axis. After about 100 iterations the difference between two successive profiles becomes imperceptible. This leads to the erroneous impression that convergence has been reached.

We have repeated the tests described in the beginning of this section, now adding Gaussian noise with a signal-to-noise ratio of 50 in the composite spectra, and random uncertainties within 1, 2, and 3 $\sigma$ in the radial velocities. This changes neither the characteristics of the spurious wings nor the convergence speed of the algorithm.

The two effects (spurious wings and incorrect line fluxes) can readily be explained. Indeed the first step of the algorithm consists of shifting the original composite spectra and add them all. This is taken as the first approximation of the spectrum of star $A$. However it should be emphasised that this step spreads the flux from star $B$ over a range of wavelengths, but does not subtract it. It is clear then that, in the given wavelength window, the total flux from star $B$ (integrated over the whole profile) is assigned entirely to star $A$:

\begin{eqnarray}
F(A_1) = F(A_0) + F(B_0)
\end{eqnarray}
where the subindex $0$ refers to the original spectra and the subindex $1$ to the first approximation.

In the second step of the first iteration, the original composite spectra are shifted and centered around the rest wavelength of star $B$, then the first approximation of star $A$ is subtracted, with the aim of enhancing the profile of star $B$. However in this operation it is the total line flux from stars $A$ and $B$ that is actually subtracted, leaving the reconstructed line profile of star $B$ with a total flux equal to zero: 

\begin{eqnarray}
 F(B_1) = F(A_0) + F(B_0) - F(A_1) = 0 
\end{eqnarray}

At each iteration, the profile of the line core of star $A$ slightly improves and the spurious absorption wings are increasingly spread out on both sides (Figure~\ref{fig:Star_1_20_40_60_80_100_200_2000}). However the integrated flux of the line of star $A$ is unchanged, and remains equal to the total from both stars. For star $B$ too the profile becomes more similar to the true one, but its total line flux remains zero, as the flux in the absorption core is equal to that of the spurious emission wings. 

After 100 iterations the algorithm seems to converge but the flux of the line of star $A$ is still much too strong and conspicuous wings are seen in the reconstructed spectra of both stars. Naturally, the stronger the flux in the line of star $B$ compared to star $A$, the stronger these effects. In this simple case, the correct profile with the correct flux is reconstructed after 2,000 iterations.

\begin{figure*}
\centerline{\includegraphics[width=2\columnwidth]{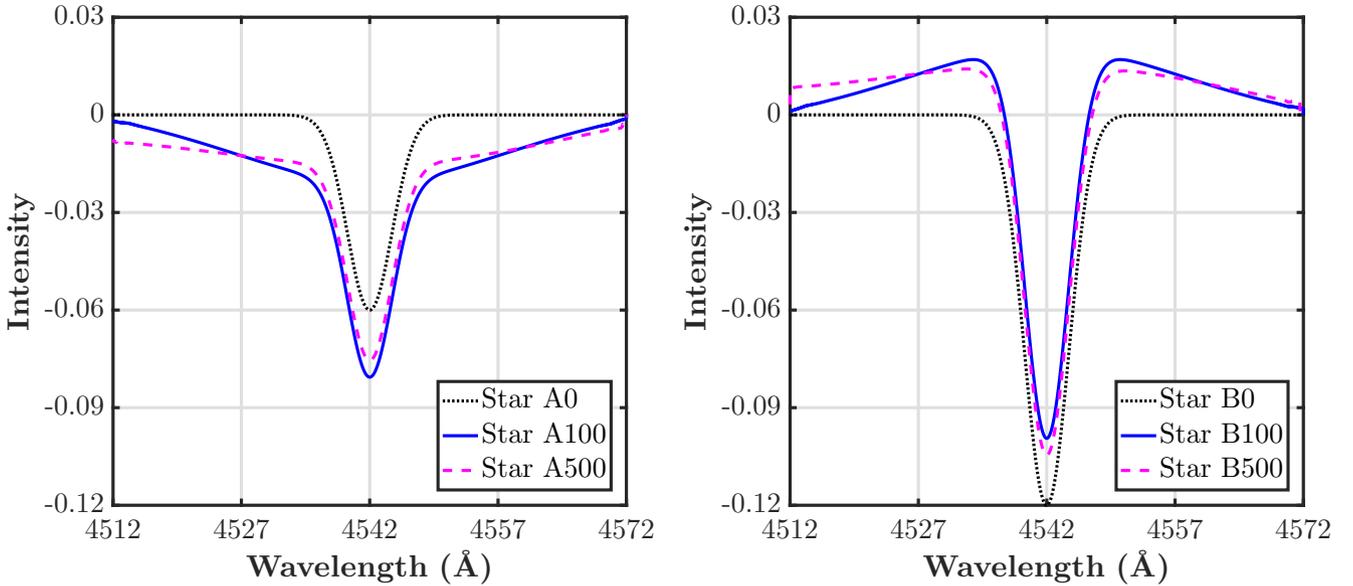}}
\caption{When applied to Gaussian profiles (dotted black lines), the MME98 algorithm generates strong spurious wings and discrepant line fluxes. 
The continuous blue and broken red lines correspond to 100 and 500 iterations respectively. Left panel, star $A$; right panel, star $B$.
\label{fig:StarA0A5B0B5Gauss}}
\end{figure*}

We also tested MME98 with simple Gaussian absorption profiles, meant to simulate roughly the He\,{\sc ii} 4542 \AA\ line of the individual components of the massive binary 9~Sgr. We assign to the line in star $B$ (the primary) half the line flux of star $A$ (Figure~\ref{fig:StarA0A5B0B5Gauss}), as suggested by Figure 3 of~\cite{Rauw2012}. We then created 51 blends of these synthetic profiles using the values of the radial velocities derived by~\cite{Rauw2012} for the 9~Sgr system. In this test, the number of wavelength steps is 2,000. Because of this increase (compared to 400 for our geometric profiles), the MME98 algorithm now requires 10,000 iterations to reconstruct the original profiles with the correct flux. This takes 35 hours on a computer with an 8-core processor at 4.20 GHz and 24 GB of RAM. And this is for a narrow spectral window covering only one line! 

\begin{figure*}
\centerline{\includegraphics[width=1.9\columnwidth]{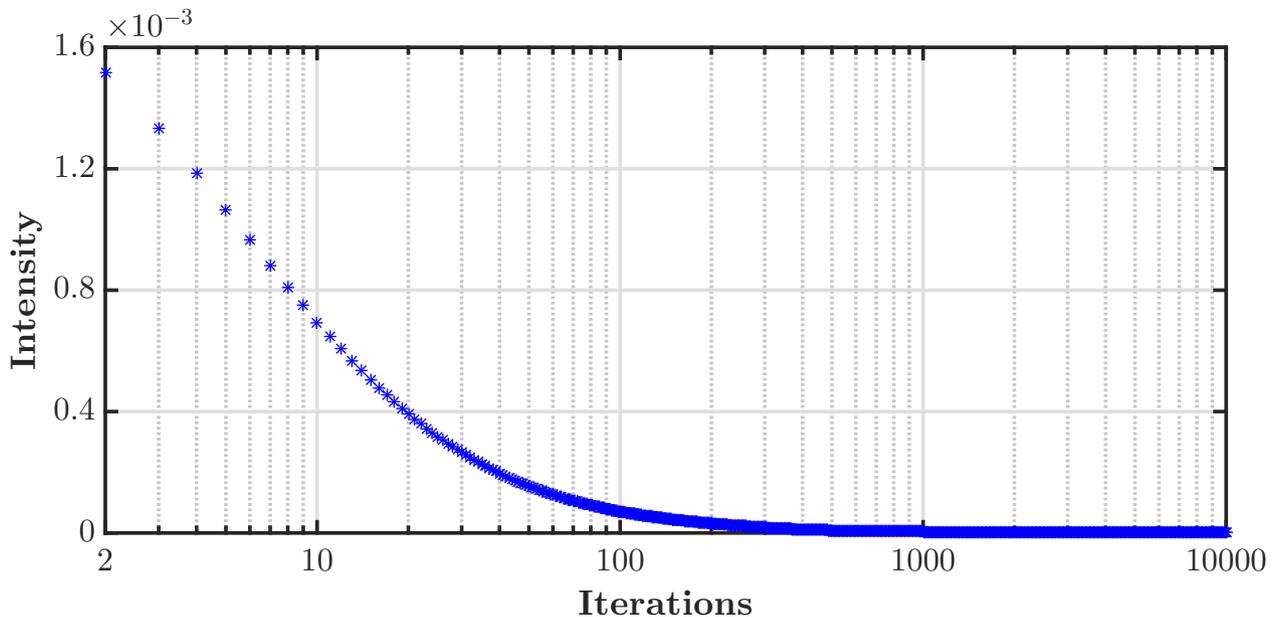}}
\caption{Root mean square of the difference between two successive reconstructions, over 10,000 iterations of the MME98 algorithm applied to composite spectra of the Gaussian profiles of Figure~\ref{fig:StarA0A5B0B5Gauss}.
\label{fig:convergencia}}  
\end{figure*}

That algorithm is therefore prohibitively slow: waiting for full convergence is practically impossible in real cases, where the spectral windows include more than one spectral line and some 12,000 wavelengths steps (as in our Figure~\ref{fig:Sgr} below) or more. We can safely conclude that most authors using MME98 never let it reach full convergence and may even believe that convergence has already been reached earlier. This illusion of convergence is due to the fact that its speed decreases exponentially (Figure~\ref{fig:convergencia}). As a consequence, there is no guarantee to obtain a faithful reconstruction by simply requesting that the difference between two successive approximations drop below the noise level or other seemingly reasonable convergence criterion. Only in experiments with synthetic input spectra can we assess the reliability of the reconstructed spectra, as we can compare them with the original profiles.

If the MME98 algorithm is stopped before full convergence, for example after 100 or 500 iterations, the disentangled profiles are similar to those obtained from the analysis of the observed spectra of 9~Sgr by~\cite{Rauw2012}, but markedly different from the original, true profiles (Figure~\ref{fig:StarA0A5B0B5Gauss}): compared with the input profiles, the reconstructed absorption line profile of star $A$ has a much larger flux and much broader absorption wings, whilst star $B$ exhibits a fainter absorption core, as well as prominent emission wings. None of these characteristics were present in the input profiles.

Adding noise to the spectra or random uncertainties to the radial velocities changes neither the speed of convergence of the MME98 algorithm nor the shape of the spurious wings. In fact, from our experiments with spectral windows of different sizes, we note that the speed of convergence depends only on the number of wavelength steps.

The limitations of the MME98 algorithm as remarked upon in the literature (see \S~\ref{sec:intro}) can be explained by these two effects: incorrect line fluxes and spurious wings. They are caused by its inability to redistribute the line flux between the two stellar components. These discrepancies are concerning, because exact line fluxes are needed to determine spectral types and, via fitting with model spectra, fundamental stellar parameters such as temperatures. Correct wings are required to characterise the surface gravity and the stellar wind.

\section{The QER20 package} \label{sec:algorithm}

We have created the QER20 package in response to the need for a highly efficient and reliable disentangling method. That is, a method that reconstructs line profiles with correct fluxes and without artefacts, and in a reasonable amount of computing time.

The disentangling methods that are most simple to use are those that work in the wavelength domain, since they do not require any space transformation. For this reason, our QER20 package is designed to work in the wavelength space, thus ensuring its efficiency.

To achieve the reliability of QER20, we have introduced a new perspective to this kind of disentangling methods: we treat the integrated spectral line flux as a free parameter. This new approach greatly accelerates the speed of convergence and thus guarantees that the output profiles are correctly reconstructed. It is worth mentioning that considering the line flux as a freely adjustable parameter during the whole process does not generate a loss of information. This is because at each iteration QER20 adjusts the reconstructed spectra to the original observations according to (~\ref{eq:starAi}) and (~\ref{eq:starBi}).

We conducted numerous experiments to probe different flux modification strategies. We describe hereafter the one that turns out to be the most efficient and reliable.

We consider first as a way of example, the case of absorption lines. Our code starts by executing the first iteration of the original MME98 method as formalised in (~\ref{eq:starA1}) and (~\ref{eq:starB1}). Next, QER20 looks for spurious emission wings in the reconstructed profile of star $B$. We do not attempt to work on the wings of the reconstructed profile of star $A$, because both its spurious wings and the real profile are in absorption, and it is not possible to separate them a priori.

To determine if emission wings in the spectrum of star $B$ are spurious, QER20 verifies that in this spectral window no emission is present in the observations. As a further test, QER20 evaluates whether these wings are symmetrically located with respect to the central wavelength of the absorption trough, since such symmetry is typical of artefacts produced by (~\ref{eq:starA1}) and (~\ref{eq:starB1}). When a spurious emission wing is identified, QER20 measures its flux and allows it to flow between the two stars during the second step (~\ref{eq:starB1}) and the two steps (~\ref{eq:starAi}) and (~\ref{eq:starBi}) of each iteration. As a result, the convergence speed dramatically accelerates and yet the correct profiles are produced, as will be shown below. This is because in our strategy, the line fluxes can be freely redistributed between the two stars.

If the lines in the observed composite spectra are in emission, our QER20 package proceeds analogously, with the spurious wings of star $B$ now in absorption. QER20 deals slightly differently if the line to disentangle is in absorption, but there is a true emission line at the position of one of the spurious emission wings, or vice-versa. (This is checked in the observed composite spectra). In that case, the flux correction strategy is applied only to the other spurious wing.

\begin{figure*}
\centerline{\includegraphics[width=2\columnwidth]{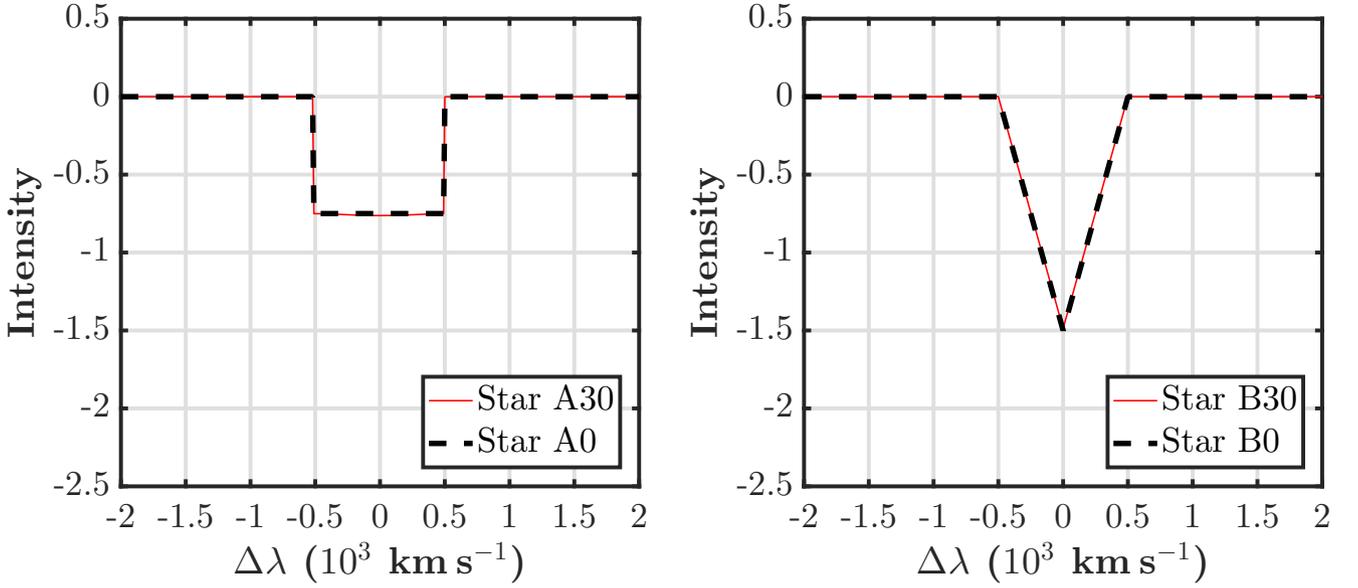}}
\caption{Output of QER20 package after 30 iterations (continuous line) compared with the original profile (dotted line). Left panel: star $A$. Right panel: star $B$.
\label{fig:StarA0A30B0B30}}  
\end{figure*}

\begin{figure*}
\centerline{\includegraphics[width=2\columnwidth]{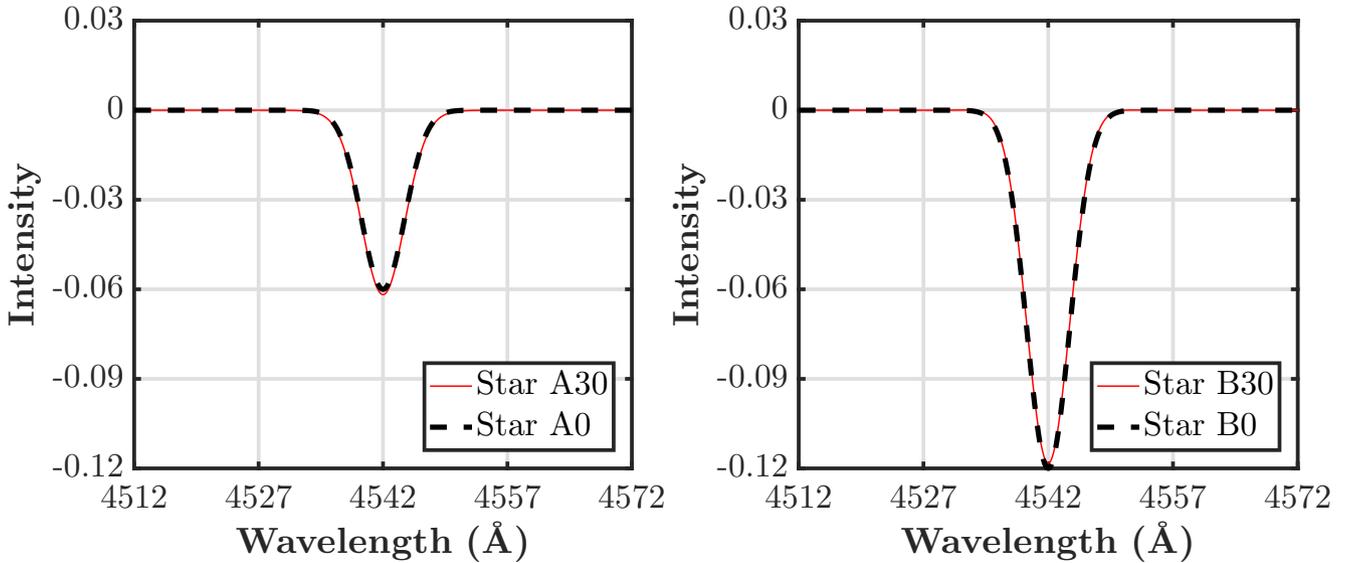}}
\caption{When applied to Gaussian profiles (dotted black lines), QER20 package produces the correct profiles after 30 iterations (continuous red lines). Left panel, star $A$; right panel, star $B$.
\label{fig:StarA0A4B0B4Gauss}}
\end{figure*}

Figure~\ref{fig:StarA0A30B0B30} illustrates the case for rectangular and triangular profiles. The convergence is reached after less than 30 iterations and the true profiles are correctly reconstructed. This result contrasts sharply with the 2,000 iterations required by the MME algorithm. In the case of Gaussian profiles, our QER20 algorithm produces the correct profiles without the strong wings and with the correct line fluxes after 30 iterations (Figure~\ref{fig:StarA0A4B0B4Gauss}), while MME requires 10,000.

As in \S~\ref{sec:analysis}, we also performed tests where Gaussian noise is added to the blends (of both geometric and Gaussian profiles) with a signal-to-noise ratio of 50, as well as random uncertainties in the radial velocities within 1, 2, and 3 $\sigma$. Under these testing conditions, the convergence speed is unchanged and the correct profiles and fluxes are obtained again, thus confirming the efficiency and the reliability of our approach.

We then apply QER20 to data simulating real conditions more closely: noisy spectra with radial velocities which are unevenly spaced and incommensurate to the wavelength step of the spectra. 

To do so we take as input the output of our disentangled spectra of 9~Sgr based on observations by~\cite{Rauw2012}, as described in the next section (\S~\ref{sec:sgr}). We construct composite spectra adding random errors within 3 $\sigma$ to the radial velocities derived by~\cite{Rauw2012}, and we add Gaussian white noise with a SNR of 50 to simulate the observed spectra of this binary. The latter step is needed because in the disentangled spectra the noise has been significantly reduced due to the averaging inherent to the method.

\begin{figure*}
\centerline{\includegraphics[width=1.8\columnwidth]{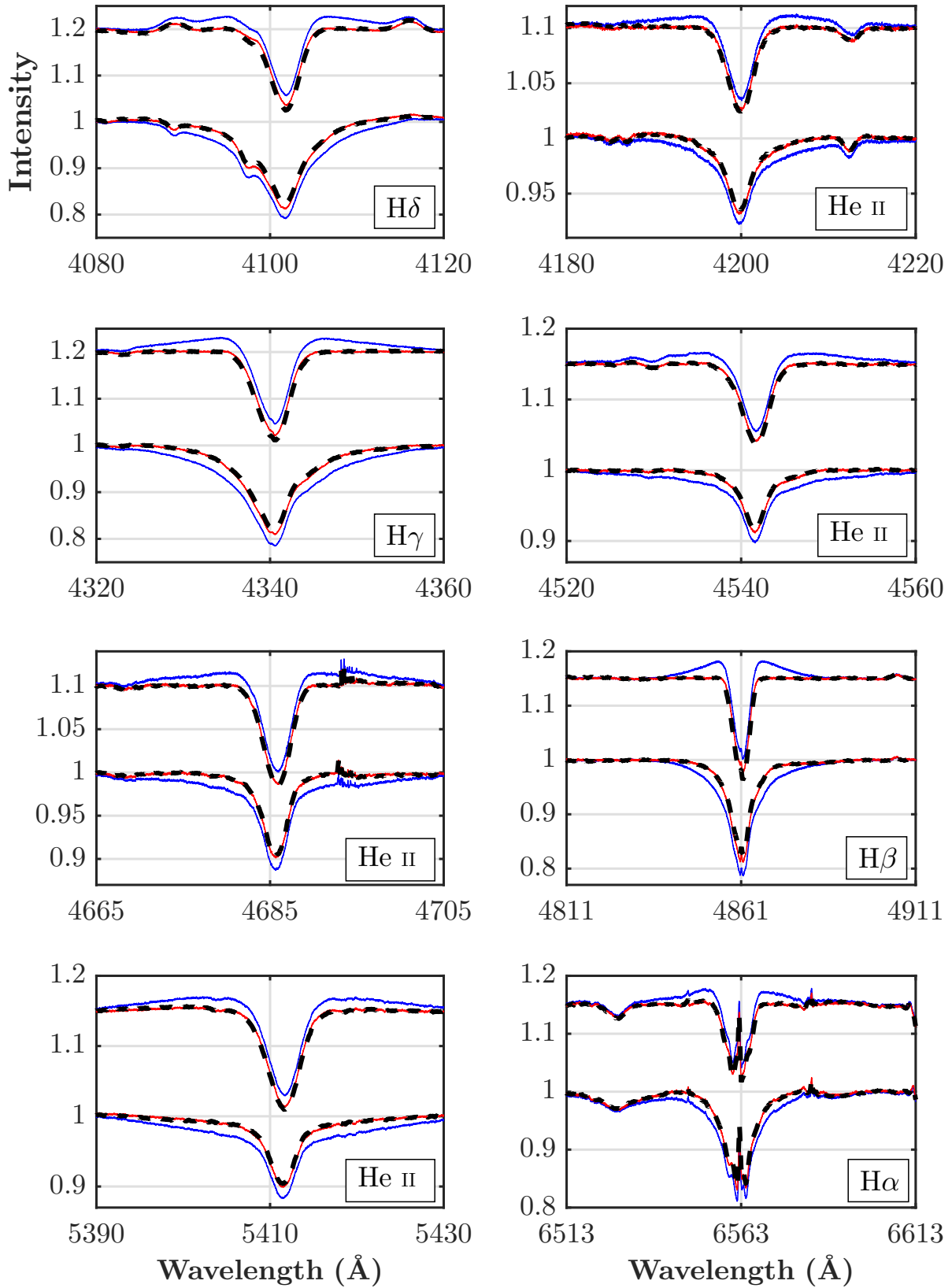}}
\caption{Comparison of the output of the MME98 after 100 iterations and QER20 after 40 iterations, for spectra simulating real conditions. Black: original spectra. Blue: spectra reconstructed with MME98. Red: spectra reconstructed with QER20. The spectra of star $B$ have been shifted vertically for clarity.
\label{fig:OriVsQEVsMME}}
\end{figure*}

After only 40 iterations our QER20 algorithm reproduces exactly the input spectra, whilst the MME98 algorithm distorts the broader lines (Balmer lines and several  He {\sc ii} lines), adding spurious absorption wings in $A$ and emission wings in $B$ (see Figure~\ref{fig:OriVsQEVsMME}).

These experiments show that our new disentangling package QER20 can reproduce exactly the line profiles and fluxes of the individual spectra, and does it in a very small number of iterations, thus fulfilling our aim of reliability and efficiency.

\section{Disentangling the Spectra of 9 Sgr}\label{sec:sgr}

Being now rather confident that our QER20 package produces disentangled spectra quite similar to those of the true individual stars, we apply it to the observations of 9~Sgr used in~\cite{Rauw2012}. This set of 51 spectra was taken with the FEROS spectrograph at La Silla Observatory with a spectral resolving power of 48\,000. The non-thermal radio emitter 9~Sgr is a massive binary consisting of two early-type main-sequence O stars orbiting each other with a period of 9.1 year and an eccentricity of 0.7 \citep{Rauw2012,Rauw2016}. 

\begin{figure*}\centering
\centerline{\includegraphics[width=2\columnwidth]{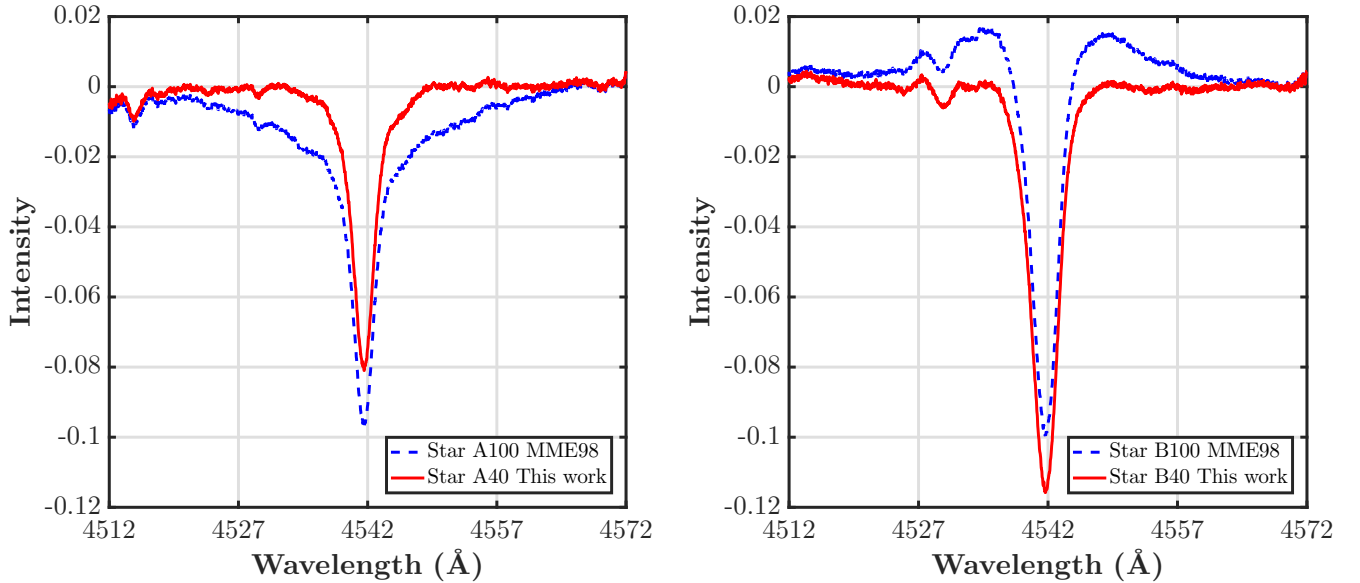}}
\caption{Reconstruction of the He\,{\sc ii} 4542 \AA\ line of 9~Sgr with our QER20 package (continuous red lines) after 40 iterations. Left panel, star $A$; right panel, star $B$. For comparison, the output of the MME98 algorithm after 100 iterations 
is shown in dotted blue lines.
\label{fig:StarA40B40Sgr}}
\end{figure*}

For the sake of comparison we first disentangle these spectra with the MME98 algorithm around the wavelength of the He\,{\sc ii} 4542 \AA\ line. The output after 100 iterations is shown in Figure~\ref{fig:StarA40B40Sgr} (dotted blue lines). The most striking features are the broad wings, in absorption in star A (left panel, dotted blue line) and in emission in star $B$ (right panel), as also seen in Figure 3 of \citet{Rauw2012}. The absence of emission in the raw data led these authors to suspect that these wings were artefacts. We confirm these suspicions since the spurious wings show high symmetry with respect to the central wavelength of the intense absorption lines.

Our QER20 package converges after only 40 iterations. The results are also displayed in Figure~\ref{fig:StarA40B40Sgr} (continuous red lines). The spurious emission and absorption wings have disappeared. We also note that the absorption line flux has significantly changed. This quantity is now larger for star $B$ whilst the opposite was the case with the MME98 algorithm. 

These differences between the output of that algorithm and our package are similar to those of our previous experiments. Because the comparison with the original profiles enabled us to confirm the exactness of the profiles reconstructed with our algorithm, we believe that we can also be fully confident in our new disentangled spectra of 9~Sgr. 

\begin{figure*}
\centerline{\includegraphics[width=1.65\columnwidth]{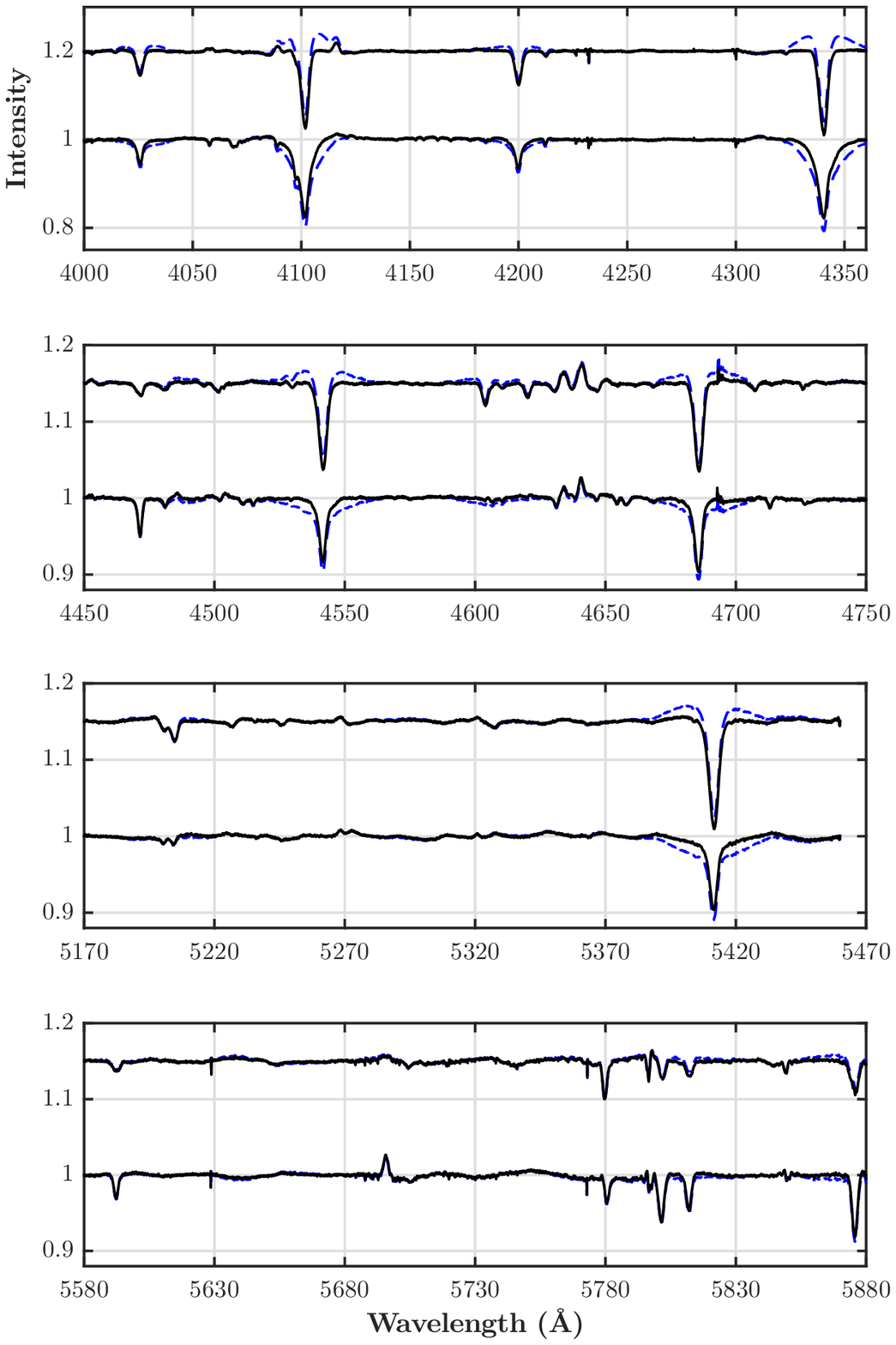}}
\caption{Disentangled spectra of 9~Sgr in the 4000 - 4360, 4450 - 4750, 5170 - 5470 and 5580 - 5880 \AA\ regions. The output of the MME98 algorithm after 100 iterations is represented with dotted blue lines, our package after 40 iterations with black continuous lines. The spectrum of the primary (star $B$) has been shifted vertically for clarity.\label{fig:Sgr}}
\end{figure*}

\begin{figure*}\centering
\centerline{\includegraphics[width=2\columnwidth]{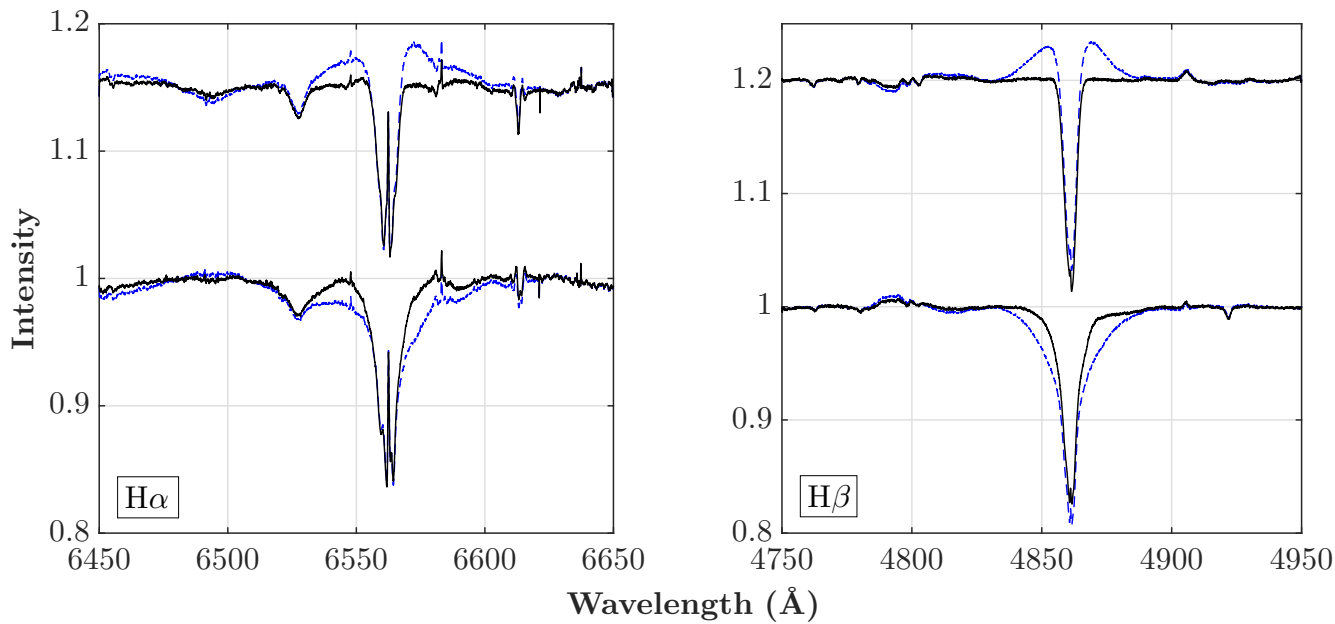}}
\caption{Disentangled spectra of 9~Sgr in the H\,$\alpha$, H\,$\beta$ regions. The output of the MME98 algorithm after 100 iterations is represented with dotted blue lines, our package after 40 iterations with black continuous lines. The spectrum of the primary (star $B$) has been shifted vertically for clarity. Several narrow nebular emission lines from the Lagoon Nebula are seen in the H\,$\alpha$ spectral region.
\label{fig:HalphaHbeta}}
\end{figure*}

The spectrum of 9~Sgr reconstructed with our QER20 package is shown in Figure~\ref{fig:Sgr} (covering most of the 4000 -- 5880 \AA\ range) and Figure~\ref{fig:HalphaHbeta} (H\,$\alpha$ and H\,$\beta$ lines). The output of the MME98 algorithm is also plotted for comparison. 

Table~\ref{tab:flux} gives the fluxes of the broadest spectral lines (Balmer lines and some He~{\sc i} and He~{\sc ii} lines). They have been measured with respect to the combined continua, as the respective contribution of each stellar component to the continuum is unknown (because disentangling techniques do not yield the stellar flux ratio). Compared to the output of the MME98 algorithm, the fluxes of the tabulated lines obtained with our package are stronger in the reconstructed spectrum of the primary star (treated as star $B$), and weaker in the secondary (star $A$). The differences are quite significant for the broadest lines because they are severely blended and thus particularly difficult to disentangle with the MME98 algorithm.  

We stress that our new package applies the flux corrections only to lines of the spectrum of star $B$ that are considered spurious, i.e. where pairs of roughly symmetric emission wings flank an absorption line. Other lines in its spectrum, such as the C\,{\sc iii} 5696 \AA\ line (Figure~\ref{fig:Sgr}), have been treated as in MME98.

\begin{center}
\begin{table*}[t]
\caption{Absorption line fluxes for 9~S\lowercase{gr}.
\label{tab:flux}} 
\centering
\begin{tabular*}{500pt}{@{\extracolsep\fill}lcD{.}{.}{3}D{.}{.}{3}cD{.}{.}{3}D{.}{.}{3}D{.}{.}{3}@{\extracolsep\fill}}
\toprule
\multicolumn{2}{c}{}
			& \multicolumn{3}{@{}c@{}}{\textbf{Secondary}}
			& \multicolumn{3}{@{}c@{}}{\textbf{Primary}}\\
  \cmidrule(r){3-5}\cmidrule(l){6-8}
  \textbf{$\lambda$} & \textbf{ID} & \multicolumn{1}{@{}l@{}}{\textbf{MME98}} & \multicolumn{1}{@{}l@{}}{\textbf{This work}} & \textbf{$\Delta$ (\%)} & \multicolumn{1}{@{}l@{}}{\textbf{MME98}} & \multicolumn{1}{@{}l@{}}{\textbf{This work}} & \multicolumn{1}{@{}l@{}}{\textbf{$\Delta$ (\%)}} \\
			\midrule
 4026  & He {\sc i}    & 56.71   & 29.43  & -48.1   & 18.9   & 26.1   & 38.2\\
 4102  & H\,$\delta$   & 257.11  & 169.78 & -34.0   & 71.0   & 98.74  & 39.1\\
 4200  & He {\sc ii}   & 72.41   & 37.26  & -48.5   & 27.25  & 36.62  & 34.4\\
 4341  & H\,$\gamma$   & 306.10  & 186.34 & -39.1   & 80.48  & 114.0  & 47.1\\
 4471  & He {\sc i}    & 14.29   & 13.79  & -3.5    & 6.50   & 6.89   & 6.0 \\ 
 4542  & He {\sc ii}   & 99.27   & 45.33  & -54.3   & 43.83  & 58.48  & 33.4\\
 4686	 & He {\sc ii}   & 84.82   & 52.31  & -38.3   & 46.84  & 57.85  & 23.5\\
 4862  & H\,$\beta$    & 336.07  & 210.10 & -37.5   & 103.25 & 128.74 & 24.7\\
 5412  & He {\sc ii}   & 149.14  & 77.14  & -48.3   & 67.16  & 84.73  & 26.2\\
 5876	 & He {\sc i}    & 31.89   & 26.69  & -16.3   & 14.63  & 19.86  & 35.7\\ 
 6563  & H\,$\alpha$   & 282.33  & 213.59 & -24.4   & 111.55 & 116.31 & 4.3\\
\bottomrule 
\end{tabular*}
\end{table*}
\end{center}

\section{Discussion and conclusions}\label{sec:conclusions}

It is quite common for massive stars to have their spectral lines significantly broadened by fast rotation and strong winds. It can then be challenging to disentangle their spectra if they belong to binary systems with low orbital velocity. We have started a series of studies with the aim to better understand the behavior of disentangling techniques. To do so, we work with blends of known individual spectra, to which the output of the algorithm can be compared. 

In this work we focus on the efficiency and reliability of the disentangling process. We have shown that the convergence of the MME98 method is in some cases too slow, and that it does not always reproduce the line fluxes and the profiles correctly. This leads to potentially misleading conclusions about the stellar type, temperature, gravity, wind, and other physical properties of the stars in the binary systems under study. Once we identified the causes of the deficiencies of this method, we were able to design a reliable and efficient method, implemented in the QER20 package, which reconstructs the individual spectra without artefacts and without error in the line fluxes, and does so in a very short computational time.

As mentioned earlier, the speed of convergence of the MME98 algorithm decreases exponentially. In very simple cases and in conditions far from real (such as our experiments with rectangular and triangular profiles), the MME98 method eventually converges to correct profiles after thousands of iterations. However, in real conditions the convergence can be so slow as to become computationally out of reach. Unfortunately, if full convergence is not reached, the MME98 output yields incorrect fluxes and incorrect wing profiles. This raises concerns about the use of the MME algorithm, either in its original form or as modified by~\cite{Gonzalez2006}.

The disentangled spectra of the stellar components of the O+O binary 9 Sgr that we obtained with the QER20 package are sufficiently different from those reconstructed by~\cite{Rauw2012} with the method of MME98 to warrant a new appraisal of their spectral types. These authors used a qualitative method of classification, namely the visual comparison with published spectra of single O stars, which has the advantage of involving many spectral lines. They assigned a subtype O\,3.5V ((f +)) to the primary (our star $B$) and O\,5-5.5V ((f)) to the secondary (star $A$). 

The quantitative classification criteria of~\cite{Conti1971} are based on the ratio of the flux of only two lines, the He~{\sc i} 4471 \AA\ and He~{\sc ii} 4542 \AA\ lines \citep[see also][]{Martins2018}. Applied to the output of the MME98 algorithm, these criteria yield O\,4V ((f +)) for the primary and O4\,V ((f )) for the secondary. However when we substitute to these line fluxes the output of our QER20 package, the subtypes we derive  are O\,3V ((f +)) and O\,5V ((f)), respectively, and are in better agreement with the qualitative classification of~\cite{Rauw2012}. 

In the case of 9~Sgr, the effect of the deficiencies of the MME98 algorithm on the classification is not severe, because the two stellar components are of similar subtype, their spectral lines have roughly the same flux and thus their ratio is not much affected. However, when a binary is composed of massive stars with different temperatures, the errors will increase. To check this, we have run tests with composite spectra of Gaussian profiles for different cases: O3+O7, O3+O9, O5+O9, etc. We note that the spectral classification of the MME98 output can be off by up to two subtypes, compared to the synthetic input. This is turn will distort the interpretation of the spectra and of the nature of the binary system, as this could mean an error of some 5000~K for the temperature and of 20 $M_ \odot$ for the masses estimated from the spectral types ~\citep{Martins2005}. In contrast, when the QER20 package is applied to those same composite spectra, there are no differences between the classification of the disentangled spectra and the synthetic input.    

We also observe that the MME98 algorithm tends to shift the subtype of one star toward that of the other. Thus it is difficult to distinguish the case of two similar stellar components from the case where they are made similar by the disentangling method.

Our strategy of taking blends of known individual spectra as input to the codes has enabled us to rigorously analyse disentangling techniques, uncover errors and propose remedies. The success of our new QER20 package encourages us to extend the present work and to study other effects, such as that of variable line profiles.

\section*{Acknowledgments}

We are grateful for financial support from the  \fundingAgency{CONACyT-DFG} bilateral project No. \fundingNumber{278156} and from the University of Guanajuato under the Apoyo a Profesores program. We thank the referee for constructive comments that have helped to improve the paper. This research has made use of NASA's Astrophysics Data System Bibliographic Services.

\bibliography{2020_07_01_9SgrQER_AN}

\begin{thebibliography}{}

\bibitem [\protect \citeauthoryear {%
{Bagnuolo}%
\ \BBA {} {Gies}%
}{%
{Bagnuolo}%
\ \BBA {} {Gies}%
}{%
{\protect \APACyear {1991}}%
}]{%
BagnuoloJr1991}
\APACinsertmetastar {%
BagnuoloJr1991}%
\begin{APACrefauthors}%
{Bagnuolo}, J., William~G.%
\BCBT {}\ \BBA {} {Gies}, D\BPBI R.%
\end{APACrefauthors}%
\unskip\
\newblock
\APACrefYearMonthDay{1991}{{\APACmonth{07}}}{},
\newblock
\unskip
\newblock
\APACjournalVolNumPages{\apj}{376}{}{266}.
\PrintBackRefs{\CurrentBib}

\bibitem [\protect \citeauthoryear {%
{Barr{\'\i}a}%
\ \protect \BOthers {.}}{%
{Barr{\'\i}a}%
\ \protect \BOthers {.}}{%
{\protect \APACyear {2013}}%
}]{%
Barria2013}
\APACinsertmetastar {%
Barria2013}%
\begin{APACrefauthors}%
{Barr{\'\i}a}, D.%
, {Mennickent}, R\BPBI E.%
, {Schmidtobreick}, L.%
\ et al.\end{APACrefauthors}%
\unskip\
\newblock
\APACrefYearMonthDay{2013}{{\APACmonth{04}}}{},
\newblock
\unskip
\newblock
\APACjournalVolNumPages{\aap}{552}{}{A63}.
\PrintBackRefs{\CurrentBib}

\bibitem [\protect \citeauthoryear {%
{Collado}%
, {Gamen}%
\BCBL {}\ \BBA {} {Barb{\'a}}%
}{%
{Collado}%
\ \protect \BOthers {.}}{%
{\protect \APACyear {2013}}%
}]{%
Collado2013}
\APACinsertmetastar {%
Collado2013}%
\begin{APACrefauthors}%
{Collado}, A.%
, {Gamen}, R.%
\BCBL {}\ \BBA {} {Barb{\'a}}, R\BPBI H.%
\end{APACrefauthors}%
\unskip\
\newblock
\APACrefYearMonthDay{2013}{{\APACmonth{04}}}{},
\newblock
\unskip
\newblock
\APACjournalVolNumPages{\aap}{552}{}{A22}.
\PrintBackRefs{\CurrentBib}

\bibitem [\protect \citeauthoryear {%
{Conti}%
\ \BBA {} {Alschuler}%
}{%
{Conti}%
\ \BBA {} {Alschuler}%
}{%
{\protect \APACyear {1971}}%
}]{%
Conti1971}
\APACinsertmetastar {%
Conti1971}%
\begin{APACrefauthors}%
{Conti}, P\BPBI S.%
\BCBT {}\ \BBA {} {Alschuler}, W\BPBI R.%
\end{APACrefauthors}%
\unskip\
\newblock
\APACrefYearMonthDay{1971}{{\APACmonth{12}}}{},
\newblock
\unskip
\newblock
\APACjournalVolNumPages{\apj}{170}{}{325}.
\PrintBackRefs{\CurrentBib}

\bibitem [\protect \citeauthoryear {%
{Demers}%
, {Moffat}%
, {Marchenko}%
, {Gayley}%
\BCBL {}\ \BBA {} {Morel}%
}{%
{Demers}%
\ \protect \BOthers {.}}{%
{\protect \APACyear {2002}}%
}]{%
Demers2002}
\APACinsertmetastar {%
Demers2002}%
\begin{APACrefauthors}%
{Demers}, H.%
, {Moffat}, A\BPBI F\BPBI J.%
, {Marchenko}, S\BPBI V.%
, {Gayley}, K\BPBI G.%
\BCBL {}\ \BBA {} {Morel}, T.%
\end{APACrefauthors}%
\unskip\
\newblock
\APACrefYearMonthDay{2002}{{\APACmonth{09}}}{},
\newblock
\unskip
\newblock
\APACjournalVolNumPages{\apj}{577}{1}{409-421}.
\PrintBackRefs{\CurrentBib}

\bibitem [\protect \citeauthoryear {%
{Eversberg}%
, {Moffat}%
\BCBL {}\ \BBA {} {Marchenko}%
}{%
{Eversberg}%
\ \protect \BOthers {.}}{%
{\protect \APACyear {1999}}%
}]{%
Eversberg1999}
\APACinsertmetastar {%
Eversberg1999}%
\begin{APACrefauthors}%
{Eversberg}, T.%
, {Moffat}, A\BPBI F\BPBI J.%
\BCBL {}\ \BBA {} {Marchenko}, S\BPBI V.%
\end{APACrefauthors}%
\unskip\
\newblock
\APACrefYearMonthDay{1999}{{\APACmonth{07}}}{},
\newblock
\unskip
\newblock
\APACjournalVolNumPages{\pasp}{111}{761}{861-870}.
\PrintBackRefs{\CurrentBib}

\bibitem [\protect \citeauthoryear {%
{Ferrero}%
, {Gamen}%
, {Benvenuto}%
\BCBL {}\ \BBA {} {Fern{\'a}ndez-Laj{\'u}s}%
}{%
{Ferrero}%
\ \protect \BOthers {.}}{%
{\protect \APACyear {2013}}%
}]{%
Ferrero2013}
\APACinsertmetastar {%
Ferrero2013}%
\begin{APACrefauthors}%
{Ferrero}, G.%
, {Gamen}, R.%
, {Benvenuto}, O.%
\BCBL {}\ \BBA {} {Fern{\'a}ndez-Laj{\'u}s}, E.%
\end{APACrefauthors}%
\unskip\
\newblock
\APACrefYearMonthDay{2013}{{\APACmonth{08}}}{},
\newblock
\unskip
\newblock
\APACjournalVolNumPages{\mnras}{433}{2}{1300-1311}.
\PrintBackRefs{\CurrentBib}

\bibitem [\protect \citeauthoryear {%
{Gonz{\'a}lez}%
\ \BBA {} {Levato}%
}{%
{Gonz{\'a}lez}%
\ \BBA {} {Levato}%
}{%
{\protect \APACyear {2006}}%
}]{%
Gonzalez2006}
\APACinsertmetastar {%
Gonzalez2006}%
\begin{APACrefauthors}%
{Gonz{\'a}lez}, J\BPBI F.%
\BCBT {}\ \BBA {} {Levato}, H.%
\end{APACrefauthors}%
\unskip\
\newblock
\APACrefYearMonthDay{2006}{{\APACmonth{03}}}{},
\newblock
\unskip
\newblock
\APACjournalVolNumPages{\aap}{448}{1}{283-292}.
\PrintBackRefs{\CurrentBib}

\bibitem [\protect \citeauthoryear {%
{Hadrava}%
}{%
{Hadrava}%
}{%
{\protect \APACyear {1995}}%
}]{%
Hadrava1995}
\APACinsertmetastar {%
Hadrava1995}%
\begin{APACrefauthors}%
{Hadrava}, P.%
\end{APACrefauthors}%
\unskip\
\newblock
\APACrefYearMonthDay{1995}{{\APACmonth{12}}}{},
\newblock
\unskip
\newblock
\APACjournalVolNumPages{\aaps}{114}{}{393}.
\PrintBackRefs{\CurrentBib}

\bibitem [\protect \citeauthoryear {%
{Hadrava}%
}{%
{Hadrava}%
}{%
{\protect \APACyear {2009}}%
}]{%
Hadrava2009}
\APACinsertmetastar {%
Hadrava2009}%
\begin{APACrefauthors}%
{Hadrava}, P.%
\end{APACrefauthors}%
\unskip\
\newblock
\APACrefYearMonthDay{2009}{{\APACmonth{01}}}{},
\newblock
\unskip
\newblock
\APACjournalVolNumPages{\aap}{494}{1}{399-402}.
\PrintBackRefs{\CurrentBib}

\bibitem [\protect \citeauthoryear {%
{Mahy}%
\ \protect \BOthers {.}}{%
{Mahy}%
\ \protect \BOthers {.}}{%
{\protect \APACyear {2012}}%
}]{%
Mahy2012}
\APACinsertmetastar {%
Mahy2012}%
\begin{APACrefauthors}%
{Mahy}, L.%
, {Gosset}, E.%
, {Sana}, H.%
, {Damerdji}, Y.%
, {De Becker}, M.%
, {Rauw}, G.%
\BCBL {}\ \BBA {} {Nitschelm}, C.%
\end{APACrefauthors}%
\unskip\
\newblock
\APACrefYearMonthDay{2012}{{\APACmonth{04}}}{},
\newblock
\unskip
\newblock
\APACjournalVolNumPages{\aap}{540}{}{A97}.
\PrintBackRefs{\CurrentBib}

\bibitem [\protect \citeauthoryear {%
{Mahy}%
, {Rauw}%
, {De Becker}%
, {Eenens}%
\BCBL {}\ \BBA {} {Flores}%
}{%
{Mahy}%
\ \protect \BOthers {.}}{%
{\protect \APACyear {2015}}%
}]{%
Mahy2015}
\APACinsertmetastar {%
Mahy2015}%
\begin{APACrefauthors}%
{Mahy}, L.%
, {Rauw}, G.%
, {De Becker}, M.%
, {Eenens}, P.%
\BCBL {}\ \BBA {} {Flores}, C\BPBI A.%
\end{APACrefauthors}%
\unskip\
\newblock
\APACrefYearMonthDay{2015}{{\APACmonth{05}}}{},
\newblock
\unskip
\newblock
\APACjournalVolNumPages{\aap}{577}{}{A23}.
\PrintBackRefs{\CurrentBib}

\bibitem [\protect \citeauthoryear {%
{Marchenko}%
, {Moffat}%
\BCBL {}\ \BBA {} {Eenens}%
}{%
{Marchenko}%
\ \protect \BOthers {.}}{%
{\protect \APACyear {1998}}%
}]{%
Marchenko1998}
\APACinsertmetastar {%
Marchenko1998}%
\begin{APACrefauthors}%
{Marchenko}, S\BPBI V.%
, {Moffat}, A\BPBI F\BPBI J.%
\BCBL {}\ \BBA {} {Eenens}, P\BPBI R\BPBI J.%
\end{APACrefauthors}%
\unskip\
\newblock
\APACrefYearMonthDay{1998}{{\APACmonth{12}}}{},
\newblock
\unskip
\newblock
\APACjournalVolNumPages{\pasp}{110}{754}{1416-1422}.
\PrintBackRefs{\CurrentBib}

\bibitem [\protect \citeauthoryear {%
{Martins}%
}{%
{Martins}%
}{%
{\protect \APACyear {2018}}%
}]{%
Martins2018}
\APACinsertmetastar {%
Martins2018}%
\begin{APACrefauthors}%
{Martins}, F.%
\end{APACrefauthors}%
\unskip\
\newblock
\APACrefYearMonthDay{2018}{{\APACmonth{08}}}{},
\newblock
\unskip
\newblock
\APACjournalVolNumPages{\aap}{616}{}{A135}.
\PrintBackRefs{\CurrentBib}

\bibitem [\protect \citeauthoryear {%
{Martins}%
, {Schaerer}%
\BCBL {}\ \BBA {} {Hillier}%
}{%
{Martins}%
\ \protect \BOthers {.}}{%
{\protect \APACyear {2005}}%
}]{%
Martins2005}
\APACinsertmetastar {%
Martins2005}%
\begin{APACrefauthors}%
{Martins}, F.%
, {Schaerer}, D.%
\BCBL {}\ \BBA {} {Hillier}, D\BPBI J.%
\end{APACrefauthors}%
\unskip\
\newblock
\APACrefYearMonthDay{2005}{{\APACmonth{06}}}{},
\newblock
\unskip
\newblock
\APACjournalVolNumPages{\aap}{436}{3}{1049-1065}.
\PrintBackRefs{\CurrentBib}

\bibitem [\protect \citeauthoryear {%
{Munoz}%
\ \protect \BOthers {.}}{%
{Munoz}%
\ \protect \BOthers {.}}{%
{\protect \APACyear {2017}}%
}]{%
Munoz2016}
\APACinsertmetastar {%
Munoz2016}%
\begin{APACrefauthors}%
{Munoz}, M.%
, {Moffat}, A\BPBI F\BPBI J.%
, {Hill}, G\BPBI M.%
\ et al.\end{APACrefauthors}%
\unskip\
\newblock
\APACrefYearMonthDay{2017}{{\APACmonth{05}}}{},
\newblock
\unskip
\newblock
\APACjournalVolNumPages{\mnras}{467}{3}{3105-3121}.
\PrintBackRefs{\CurrentBib}

\bibitem [\protect \citeauthoryear {%
{Pavlovski}%
\ \BBA {} {Hensberge}%
}{%
{Pavlovski}%
\ \BBA {} {Hensberge}%
}{%
{\protect \APACyear {2010}}%
}]{%
Pavlovski2010}
\APACinsertmetastar {%
Pavlovski2010}%
\begin{APACrefauthors}%
{Pavlovski}, K.%
\BCBT {}\ \BBA {} {Hensberge}, H.%
\end{APACrefauthors}%
\unskip\
\newblock
\APACrefYearMonthDay{2010}{{\APACmonth{12}}}{},
\newblock
{\BBOQ}\APACrefatitle {{Reconstruction and Analysis of Component Spectra of
  Binary and Multiple Stars}} {{Reconstruction and Analysis of Component
  Spectra of Binary and Multiple Stars}}.{\BBCQ}
\newblock
\BIn{} A.~{Pr{\v{s}}a}\ \BBA {} M.~{Zejda}\ (\BEDS), \APACrefbtitle {Binaries -
  Key to Comprehension of the Universe} {Binaries - Key to Comprehension of the
  Universe}\ \BVOL~435, \BPG~207.
\PrintBackRefs{\CurrentBib}

\bibitem [\protect \citeauthoryear {%
{Raucq}%
\ \protect \BOthers {.}}{%
{Raucq}%
\ \protect \BOthers {.}}{%
{\protect \APACyear {2016}}%
}]{%
Raucq2016}
\APACinsertmetastar {%
Raucq2016}%
\begin{APACrefauthors}%
{Raucq}, F.%
, {Rauw}, G.%
, {Gosset}, E.%
, {Naz{\'e}}, Y.%
, {Mahy}, L.%
, {Herv{\'e}}, A.%
\BCBL {}\ \BBA {} {Martins}, F.%
\end{APACrefauthors}%
\unskip\
\newblock
\APACrefYearMonthDay{2016}{{\APACmonth{04}}}{},
\newblock
\unskip
\newblock
\APACjournalVolNumPages{\aap}{588}{}{A10}.
\PrintBackRefs{\CurrentBib}

\bibitem [\protect \citeauthoryear {%
{Rauw}%
\ \protect \BOthers {.}}{%
{Rauw}%
\ \protect \BOthers {.}}{%
{\protect \APACyear {2016}}%
}]{%
Rauw2016}
\APACinsertmetastar {%
Rauw2016}%
\begin{APACrefauthors}%
{Rauw}, G.%
, {Blomme}, R.%
, {Naz{\'e}}, Y.%
\ et al.\end{APACrefauthors}%
\unskip\
\newblock
\APACrefYearMonthDay{2016}{{\APACmonth{05}}}{},
\newblock
\unskip
\newblock
\APACjournalVolNumPages{\aap}{589}{}{A121}.
\PrintBackRefs{\CurrentBib}

\bibitem [\protect \citeauthoryear {%
{Rauw}%
\ \protect \BOthers {.}}{%
{Rauw}%
\ \protect \BOthers {.}}{%
{\protect \APACyear {2012}}%
}]{%
Rauw2012}
\APACinsertmetastar {%
Rauw2012}%
\begin{APACrefauthors}%
{Rauw}, G.%
, {Sana}, H.%
, {Spano}, M.%
, {Gosset}, E.%
, {Mahy}, L.%
, {De Becker}, M.%
\BCBL {}\ \BBA {} {Eenens}, P.%
\end{APACrefauthors}%
\unskip\
\newblock
\APACrefYearMonthDay{2012}{{\APACmonth{06}}}{},
\newblock
\unskip
\newblock
\APACjournalVolNumPages{\aap}{542}{}{A95}.
\PrintBackRefs{\CurrentBib}

\bibitem [\protect \citeauthoryear {%
{Schnurr}%
, {Moffat}%
, {Villar-Sbaffi}%
, {St-Louis}%
\BCBL {}\ \BBA {} {Morrell}%
}{%
{Schnurr}%
\ \protect \BOthers {.}}{%
{\protect \APACyear {2009}}%
}]{%
Schnurr2009}
\APACinsertmetastar {%
Schnurr2009}%
\begin{APACrefauthors}%
{Schnurr}, O.%
, {Moffat}, A\BPBI F\BPBI J.%
, {Villar-Sbaffi}, A.%
, {St-Louis}, N.%
\BCBL {}\ \BBA {} {Morrell}, N\BPBI I.%
\end{APACrefauthors}%
\unskip\
\newblock
\APACrefYearMonthDay{2009}{{\APACmonth{05}}}{},
\newblock
\unskip
\newblock
\APACjournalVolNumPages{\mnras}{395}{2}{823-836}.
\PrintBackRefs{\CurrentBib}

\bibitem [\protect \citeauthoryear {%
{Shenar}%
\ \protect \BOthers {.}}{%
{Shenar}%
\ \protect \BOthers {.}}{%
{\protect \APACyear {2018}}%
}]{%
Shenar2018}
\APACinsertmetastar {%
Shenar2018}%
\begin{APACrefauthors}%
{Shenar}, T.%
, {Hainich}, R.%
, {Todt}, H.%
\ et al.\end{APACrefauthors}%
\unskip\
\newblock
\APACrefYearMonthDay{2018}{{\APACmonth{08}}}{},
\newblock
\unskip
\newblock
\APACjournalVolNumPages{\aap}{616}{}{A103}.
\PrintBackRefs{\CurrentBib}

\bibitem [\protect \citeauthoryear {%
{Tramper}%
\ \protect \BOthers {.}}{%
{Tramper}%
\ \protect \BOthers {.}}{%
{\protect \APACyear {2016}}%
}]{%
Tramper2016}
\APACinsertmetastar {%
Tramper2016}%
\begin{APACrefauthors}%
{Tramper}, F.%
, {Sana}, H.%
, {Fitzsimons}, N\BPBI E.%
, {de Koter}, A.%
, {Kaper}, L.%
, {Mahy}, L.%
\BCBL {}\ \BBA {} {Moffat}, A.%
\end{APACrefauthors}%
\unskip\
\newblock
\APACrefYearMonthDay{2016}{{\APACmonth{01}}}{},
\newblock
\unskip
\newblock
\APACjournalVolNumPages{\mnras}{455}{2}{1275-1281}.
\PrintBackRefs{\CurrentBib}

\end{thebibliography}

\end{document}